\documentclass{article}     
\catcode`\@=11
\@addtoreset{equation}{section}

\def\@normalsize{\@setsize\normalsize{15pt}\xiipt\@xiipt
\abovedisplayskip 14pt plus3pt minus3pt%
\belowdisplayskip \abovedisplayskip
\abovedisplayshortskip \z@ plus3pt%
\belowdisplayshortskip 7pt plus3.5pt minus0pt}

\def\small{\@setsize\small{13.6pt}\xipt\@xipt
\abovedisplayskip 13pt plus3pt minus3pt%
\belowdisplayskip \abovedisplayskip
\abovedisplayshortskip \z@ plus3pt%
\belowdisplayshortskip 7pt plus3.5pt minus0pt
\def\@listi{\parsep 4.5pt plus 2pt minus 1pt
     \itemsep \parsep
     \topsep 9pt plus 3pt minus 3pt}}

\relax
\catcode`@=12
\catcode`\@=11
\def\section{\@startsection{section}{1}{\z@}{3.5ex plus 1ex minus
   .2ex}{2.3ex plus .2ex}{\large\bf}}
\def\thesection{\arabic{section}.}

\def\appendix{\setcounter{section}{0}
 \def\thesection{Appendix \Alph{section}:}
 \def\theequation{\Alph{section}.\arabic{equation}}}

\begin{document}
       
\begin{titlepage}
\begin{center}
{\Large   Dynamics  of          
  Supersymmetric  $SU(n_c)$  and $USp(2 n_c)$ Gauge 
Theories   }
\end{center}

\vspace{1em}
\begin{center}
{\large    Giuseppe Carlino$^{(1,2,3)}$, Kenichi Konishi$^{(4,5,6)}$ \\  
and \\ Hitoshi Murayama$^{(7,8) }$ }
\end{center}
\vspace{1em}
\begin{center}
{\it {
$^{(1)}$Dipartimento di Fisica,  Universit\`a di Genova   and     
$^{(2)}$Istituto Nazionale di Fisica Nucleare,    \\  Sezione di Genova,    
Via Dodecaneso, 33 -- 16146 Genova (Italy); \\
$^{(3)}$Department of Physics, University of Wales Swansea, \\Singleton Park,
Swansea SA2 8PP, United Kingdom\\
$^{(4)}$Dipartimento di Fisica, Universit\`a di Pisa   and   
$^{(5)}$Istituto Nazionale di Fisica Nucleare,   \\    Sezione di Pisa, 
Via Buonarroti, 2, Ed.B-- 56127 Pisa (Italy)\\
$^{(6)}$Department of Physics,  University of Washington, \,\,    
Seattle,   WA 19185 (USA)   \\  
$^{(7)}$Department of Physics,
University of California, Berkeley, CA 94720 (USA) \\
$^{(8)}$Theoretical Physics Group, Lawrence Berkeley National Laboratory\\
1 Cyclotron Road, Berkeley, CA 94720 (USA)\\
E-mail: {\tt carlino@ge.infn.it  konishi@phys.washington.edu  murayama@lbl.gov}
}   }  
\end{center}     
\vspace{3em}
\noindent
{\bf ABSTRACT:}   {   We study dynamical flavor symmetry breaking  in the context of a class of $N=1$
  supersymmetric $SU(n_c)$ and $USp(2 n_c)$ gauge theories,   constructed  
 from the exactly solvable $N=2$
  theories    by perturbing them with  small adjoint and generic bare
  hypermultiplet (quark) masses.     We find that the flavor $U(n_{f})$
  symmetry in $SU(n_{c})$ theories is dynamically broken to $U(r)
  \times U(n_{f}-r)$ groups for $n_f \leq n_c$.  In the $r=1$ case
the dynamical symmetry breaking is caused  by the condensation of monopoles in the $\underline{n_{f}}$
  representation.  For general $r$, however, the monopoles in the
  $\underline{{}_{n_{f}}C_{r}}$ representation, whose condensation
  could  explain the flavor symmetry breaking but would  produce
  too-many Nambu--Goldstone multiplets, actually ``break up'' into
  ``magnetic quarks'' which  condense and induce confinement and the  symmetry breaking. 
 In $USp(2n_c)$ theories
  with $n_f \leq n_c + 1$, the flavor $SO(2n_f)$ symmetry is
  dynamically broken to $U(n_f)$, but with no description in terms of
  a weakly coupled local field theory.   In both $SU(n_c)$  
  and $USp(2 n_c)$ theories,  with larger numbers  of quark flavors,   besides the vacua with these properties, 
there exist
  also  vacua with no flavor symmetry breaking.}

\vspace{1.5em}
\begin{flushleft}
GEF-TH 6/99; IFUP-TH 53/99;  UCB-PTH-99/57; LBNL-44783
\end{flushleft}   

\begin{flushright}
January  2000
\end{flushright}
\end{titlepage}

\newcommand{\1}{{\Bbb I}}
\newcommand{\Z}{{\Bbb Z}}
\newcommand{\beq}{\begin{equation}}
\newcommand{\eeq}{\end{equation}}
\newcommand{\bea}{\begin{eqnarray}}
\newcommand{\eea}{\end{eqnarray}}
\newcommand{\beas}{\begin{eqnarray*}}
\newcommand{\eeas}{\end{eqnarray*}}
\newcommand{\defi}{\stackrel{\rm def}{=}}
\newcommand{\non}{\nonumber}
\def\dirac{{\cal D}}
\def\dplus{{\cal D_{+}}}
\def\dminus{{\cal D_{-}}}
\def\dbar{\bar{D}}
\def\L{{\mathcal L}}
\def\H{\cal{H}}
\def\de{\partial}
\def\si{\sigma}
\def\sb{{\bar \sigma}}
\def\rn{{\bf R}^n}
\def\r4{{\bf R}^4}
\def\s4{{\bf S}^4}
\def\ker{\hbox{\rm ker}}
\def\dim{\hbox{\rm dim}}
\def\sup{\hbox{\rm sup}}
\def\inf{\hbox{\rm inf}}
\def\infi{\infty}
\def\nrm{\parallel}
\def\nrmi{\parallel_\infty}
\def\om{\Omega}
\def\Tr{ \hbox{\rm Tr}}
\def\const{\hbox {\rm const.}}
\def\o{\over}
\def\th{\theta}
\def\im{\hbox{\rm Im}}
\def\re{\hbox{\rm Re}}
\def\bra{\langle}
\def\ket{\rangle}
\def\Arg{\hbox {\rm Arg}}
\def\Re{\hbox {\rm Re}}
\def\Im{\hbox {\rm Im}}
\def\diag{\hbox{\rm diag}}

\vspace{1.5em}

\noindent {\bf {1. } }   An interesting phenomenon has been
observed in $N=2$ supersymmetric $SU(2)$ gauge theories with various
flavors and with adjoint mass perturbation \cite{SW1,SW2}: confinement
is caused by condensation of magnetic monopoles carrying nontrivial
flavor quantum numbers (see also \cite{KT} for further details):
spontaneous flavor symmetry breaking is  caused by the same
dyamical mechanism responsible for confinement in these models.  
We wish to know what happens in more general classes of models, and
through a systematic analysis, to gain a more microscopic
understanding of these phenomena and related ones in   Quantum Chromodynamics.  As we see below, the
generalization from $SU(2)$     to higher-rank  gauge groups
turns out to be quite subtle.

We  discuss here models   constructed from
exactly solvable $N=2$ $SU(n_c)$ and $USp(2n_c)$ gauge theories with
all possible numbers of flavor compatible with asymptotic freedom, by
perturbing them with a small adjoint mass (reducing supersymmetry to
$N=1$) and keeping small, generic bare hypermultiplet (quark) masses.
The advantage of doing so is that the only vacua retained are those in
which the gauge coupling constant grows in the infrared.  Another
advantage is that in this way all flat directions are eliminated and
one is left with a finite number of isolated vacua; keeping track of
this number allows us to perform highly nontrivial checks of our
analyses at various steps.
Our analysis heavily relies on   the breakthrough works by Seiberg
and Witten \cite{SW1,SW2}, and those which followed them
\cite{SUN}.    Also   crucial    will be Seiberg's
$N=1$ electromagnetic duality \cite{Sei,Altri}, and newly discovered
universal classes of (super) conformally invariant theories
\cite{Sei}-\cite{EHIY}.

 The special cases of 
$SU(2)=USp(2)$ theories with $n_{f} = 1, 2, 3, 4$  were studied in  
\cite{SW2}.  For $n_{f}=1,4$, there is no dynamical flavor symmetry 
breaking.  For $n_{f}=2$, monopoles in the 
$(\underline{2},1)+(1,\underline{2})$ (spinor) representation of the 
flavor $[SU(2)\times SU(2)]/Z_{2}=SO(4)$ group is found to  condense after $N=1$ 
perturbation $\mu \,\Tr \, \Phi^2$: the flavor symmetry is 
necessarily broken to $U(2)$.  For $n_{f}=3$, monopoles in the 
$\underline{4}$ (spinor) representation of the flavor $SO(6)$ group 
condense with $\mu\neq 0$ and the flavor symmetry is broken to $U(3)$ 
while there is another vacuum where a flavor-singlet dyon condenses 
and the flavor symmetry is unbroken.  This result naturally leads to a 
conjecture that the condensation of monoples with non-trivial flavor 
transformation property explains the confinement {\it \`a la}\/ 
`t Hooft \cite{TH}  and the flavor symmetry breaking simultaneously.  However, a 
simple thought reveals a problem with this picture.  As we will see 
later, the monopoles in $USp(2n_{c})$ theories transform under the 
spinor representation of $SO(2n_{f})$ flavor symmetry, and their 
effective low-energy Lagrangian coupled to the magnetic $U(1)$ gauge 
group would have an accidental $SU(2^{n_{f}-1})$ flavor symmetry, and 
their condensation would lead to far too many Nambu--Goldstone 
multiplets.  The case of $SU(2)$ gauge theories was special because 
the flavor symmetries of the monopole action precisely coincide with 
the symmetry of the microscopic theories due to the small number of 
flavors.  This argument suggests that the phenomenon of flavor 
symmetry breaking is richer in higher rank theories.  

Argyres, Plesser and Seiberg \cite{ArPlSei} studied higher-rank
$SU(n_{c})$ theories with $n_{f} \le 2n_{c}-1$ (asymptotically    free) in
detail.  They showed how the non-renormalization theorem of the
hyperK\"ahler metric on the Higgs branch could  be used to show the
persistence of unbroken non-abelian gauge group at the ``roots'' of
the Higgs branches (non-baryonic and baryonic branches) where they
intersect the Coulomb branch.  Some isolated points on the
non-baryonic roots with $SU(r)$ ($r \leq [n_{f}/2]$) gauge group as
well as the baryonic root (single point) with $SU(\tilde{n}_{c}) =
SU(n_{f}-n_{c})$ gauge group were  found to survive the $\mu \neq 0$ perturbation.
Their main focus, however, was the attempt to ``derive'' Seiberg's  
duality between $SU(n_{c})$ and $SU(\tilde{n}_{c})$ gauge theories
relying on the baryonic root,\footnote{This ``derivation,'' however,
  was incomplete as it did not produce all components of the ``meson''
  superfield.  Moreover, the effective low-energy theory was perturbed
  by a relevant operator (the mass term for the mesons) and did not
  flow to the Seiberg's magnetic theory correctly.  We thank P.
  Argyres for discussions on this point.} and the issue of flavor
symmetry breaking was not studied at any depth.  The analysis also left a puzzle
why there were ``extra'' theories at the non-baryonic roots which
seemingly had nothing to do with  Seiberg's    dual theories.  Another  
paper by Argyres, Plesser and Shapere addressed  similar questions   in    $SO(n_{c})$ and $USp(2n_{c})$
theories \cite{APS2}.  

In the present    paper, we find that the flavor $U(n_{f})$ symmetry in
$SU(n_{c})$ theories can be dynamically broken to  various  $U(r) \times
U(n_{f}-r)$ groups.  We find that  in the $r=1$ vacua    the dynamical symmetry 
breaking  is indeed caused by the
condensation of monopoles in the $\underline{n_{f}}$ representation.
For general $r$, however, the monopoles in the
$\underline{{}_{n_{f}}C_{r}}$ representation, whose condensation could
have explained the flavor symmetry breaking but would   have produced too-many
Nambu--Goldstone multiplets, actually ``break up'' into ``magnetic
quarks'' whose baryonic composites under the unbroken $SU(r)$ gauge
group match the monopoles.  The baryonic roots are shown always to
coincide with the non-baryonic roots with $r=\tilde{n}_{c}$.  The
non-baryonic roots are shown to be necessary ingredients of the
Seiberg's dual theories rather than being ``extra.''  The vacua with
unbroken flavor symmetries are associated with the baryonic roots.
The situation with $USp(2n_{c})$ theories is    even less   trivial.
The low-energy theories are non-trivial superconformal theories with
no description in terms of a weakly coupled local field theory.  In
obtaining these results, counting of the number of vacua proved to be
an extremely useful tool. The counting was done in the semi-classical
limit, large $\mu$ limit, using the curve, as well as using  low-energy
effective Lagrangians and    they all agree   with each other.

\smallskip

\noindent  {\bf  2. }       First we perform a preparatory
analysis,  by minimizing the
scalar potential following from the Lagrangian valid
in the semi-classical regime (when both $\mu$ and $m$ are large).  
$N=1$ supersymmetry and holomorphy guarantee   the absence of phase 
transitions between large $\mu$, $m$ to small $\mu$, $m$.  Therefore 
these vacua are related to quantum vacua  in other regimes  one by one.  

The Lagrangian of the models has the structure
\begin{equation}
  \L=     {1\over 8 \pi} \im \, \tau_{cl} \left[\int d^4 \th \,
    \Phi^{\dagger} e^V \Phi +\int d^2 \th\,{1\o 2} W W\right]
  + \L^{(quarks)} + \Delta \L,
  \label{lagrangian}
\end{equation}
where
\begin{equation}  
  \Delta \L=   \int \, d^2 \theta \,\mu  \,\Tr \, \Phi^2
  \label{adjointmass} 
\end{equation}
is the adjoint mass breaking the supersymmetry to $N=1$ and
\begin{equation} 
  \L^{(quarks)}= \sum_i \, \left[ \int d^4 \th \, \{ Q_i^{\dagger} e^V 
    Q_i + {\tilde Q_i}  e^{-V} {\tilde Q}_i^{\dagger} \} +  \int d^2\th\, 
    \{ \sqrt{2} {\tilde Q}_i \Phi Q^i   +      m_i   {\tilde Q}_i \Phi
    Q^i   \} \right]
    \label{lagquark}   
\end{equation}
describes the $n_{f}$ flavors of hypermultiplets (``quarks''),  and
$
\tau_{cl} \equiv  {\th_0 / \pi} + {8 \pi i/g_0^2}
$
is the bare $\theta$ parameter and coupling constant.
The $N=1$ chiral and gauge superfields $\Phi= \phi \, + \, \sqrt2 \,
\th \,\psi + \, \ldots \, $, and $W_{\alpha} = -i \lambda \, + \, {i
\o 2} \, (\si^{\mu} \, \sb^{\nu})_{\alpha}^{\beta} \, F_{\mu \nu} \,
\th_{\beta} + \, \ldots $ are both in the adjoint representation of
the gauge group, while the quarks   are taken in the
fundamental representation.
In the  limit 
$m_i  \to 0,  $  and $\mu\to 0$,    these  models possess  an exact 
flavor symmetry,   $U(n_f) \times Z_{2n_c-n_f} $  or $SO(2
n_f)\times Z_{2n_c +2 - n_f}$,   for  
$SU(n_c)$ or
$USp(2n_c)$  gauge groups,     respectively.    In the equal  quark  mass limit, the symmetry of the symplectic gauge theory  is
reduced to
  $U(n_f). $
 The models
are asymptotically free as long as 
$   n_f < 2 n_c$ (for  $SU(n_c)$ gauge theory)    or $n_f <  2n_c +2$ (for
$USp(2n_c)$). 

We find
\begin{equation}
  {\cal N} = \sum_{r=0}^{{\rm min} \, \{n_f, n_c-1\}}\, (n_c-r)
  {}_{n_f}C_{r}
  \label{nofvac}
\end{equation}
semi-classical solutions for $SU(n_c)$ gauge theory with $n_f$
flavors, while for $Usp(2n_c)$ theory with $n_f$ flavors, the number
of $N=1$ vacua is
\begin{equation}
  {\cal N} = \sum_{r=0}^{{\rm min}\{ n_c, \, n_f\}}  (n_c- r +1) \,
  {}_{n_f}\!C_{r} \, .
  \label{Nvspnclass}
\end{equation}
The factor $n_c-r$ or $n_c- r +1$ appearing in the sum originates from
Witten's index for unbroken gauge group.  For small number of flavors,
these expressions simplify somewhat:
\begin{equation}
  {\cal N}_1= ( 2 \, n_c - n_f) \, 2^{n_f -1}, \qquad (SU(n_c) \,\,
  {\hbox{\rm  with}}  \,\,n_f \le n_c);  \label{nofvacbis} 
\end{equation}
\begin{equation}
  {\cal N}_1 =(2 \, n_c+2-n_f) \, 2^{n_f-1},  \qquad   
  ( USp(2n_c)   \,\, {\hbox{\rm  with}}  \,\ n_f \le n_c+1).   
  \label{Nvspnclassbis} 
\end{equation}
It is amusing that   these different expressions all   reproduce correctly
the number of $N=1$ vacua in the case of $SU(2)$ theory (which is a
special case, both of $SU(n_c)$ and of $USp(2n_c)$) with $n_f=0 \sim 
4,$
\begin{equation}    
  {\cal N}= n_f +2.  
\end{equation}

\noindent {\bf  3. }        We next       determine the possible patterns of 
dynamical flavor symmetry breaking in these theories.  This is done 
most easily by studying these theories at large fixed $\mu \gg 
\Lambda$, $m_{i} \to 0$.\footnote{We also investigated  the limit $ \mu \to 
\infty$ while $m_i \ll \Lambda $ fixed, which is suited for studying 
the decoupling of the adjoint fields.  We checked   this way  the consistency 
with the known results about $N=1$ theories.} Such an 
analysis is possible since at large adjoint mass the low-energy 
effective superpotential can be read off from the bare Lagrangian by 
integrating out the heavy adjoint field and by adding to it the 
known exact instanton--induced superpotentials of the corresponding 
$N=1$ theories.  By minimizing the superpotential, we found in all 
cases the correct number of vacua 
Eqs.(\ref{nofvac})-(\ref{Nvspnclassbis}).  $N=1$ supersymmetry kept 
intact throughout guarantees that there are no phase transitions as 
$\mu$ is varied; we can thus determine the symmetry breaking pattern 
in each $N=1$ vacuum from the first principles.  The analysis is 
straightforward, but is not entirely trivial for large $n_{f}$ because 
the non-perturbative effects among the low-energy degrees of freedom 
(dual quarks and mesons) have to be correctly taken into account 
despite the fact that they are in a ``free magnetic phase''\cite{Sei}.  

For instance,   for $SU(n_c)$ theory with  $n_f <  n_c$ the effective
superpotential reads  
\begin{equation}
  W =  -{1 \o 2 \mu} \left[ \Tr M^2 - {1 \o n_c}(\Tr M)^2 \right]  
  + \Tr (m M )  +
  {\Lambda_1^{(3n_c - n_f)/(n_c-n_f)}  \o (\det M)^{1/(n_c-n_f)}} \, ,
  \label{splarmusmnf}
\end{equation}
where $M_i^j \equiv {\tilde Q}_i^a Q_a^j$, and $ \Lambda_1 =   
(\mu^{n_{c}} \Lambda^{2 n_c-n_f})^{1 \o 3 n_c-n_f}$    is the scale of
the $N=1$ theory.     The    minima of the potential are characterized by
the set of vacuum expectation values (in the $m_i \to 0$ limit),
\begin{equation}   
  M = \diag \, (\lambda_1, \lambda_2,  \ldots, \lambda_{n_f}), 
  \label{missing} 
\end{equation}
\begin{equation}  
  \lambda_1=\ldots =\lambda_r= 
  -(n_{c}+r-n_{f}) Z, \qquad 
  \lambda_{r+1}=\ldots =\lambda_{n_f} = 
  (n_{c}-r) Z, 
\end{equation}
where 
\begin{equation}  
  Z=  C \, \left(\mu^{n_c-n_f} \Lambda_{1}^{3n_{c}-n_{f}}
  \right)^{1/ (2 n_c- n_f)} \omega^{k}, 
  \qquad  (k=1,2, \ldots
   2 n_c- n_f; \, \omega=e^{2\pi i  /(2 n_c- n_f)}),
   \label{usefuleq2} 
\end{equation}   
with $C \sim O(1)$ a constant that depends on $n_f$, $n_c$ and $r$.  
In a vacuum characterized by
$r$, the flavor symmetry of the model is broken spontaneously as
\begin{equation}  
  U(n_f) \to U(r) \times U(n_f-r).    
  \label{symmetrybr} 
\end{equation}
To avoid double counting, we can restrict $r \leq [n_{f}/2]$ with 
all $k$, and for the special case of $r=n_{f}/2$ (possible only when 
$n_{f}$ is even), $k=1, \cdots, n_{c}-n_{f}/2$, for each choice of $r$ 
flavors out of $n_{f}$.  We find a total
\begin{equation}    
  {\cal N}_1  =  ( 2 n_c- n_f ) \cdot  2^{ n_f -1}      
  \label{slnsmz}    
\end{equation}
of such vacua, after summation over $r$.  
The number for $n_f=n_c$ is given by the same formula using the
``quantum modified constraint'' among the mesons and baryons following
Seiberg.   For  $n_f \le n_c$  the above exhausts the number of the vacua.   In the case $n_f=n_c+1$ we used  the appropriate  effective
Lagrangian  involving mesons and  baryons  to find that there are  $  {\cal N}_1 $ vacua   with various symmetry breaking
(\ref{symmetrybr})  plus one vacuum with  no flavor symmetry breaking.  The total number   $  {\cal N}_1 +1 $ reproduces  (\ref{nofvac})
correctly.

      The situation    for larger  numbers of flavor    ($n_f >  n_c +1 $)
is  more subtle.      The effective low-energy action in these
cases   has the form (we set the ``matching scale'' to unity to
simplify expressions)
\begin{equation}    
   W = {\tilde q} M q  +    \Tr ( m M )   
  -{1 \o 2 \mu} \left[ \Tr M^2 - {1 \o n_c}(\Tr M)^2 \right],
  \label{dualqeq}
\end{equation}
where $q$'s are $n_f$ flavors of dual quarks \cite{Sei} in the
fundamental representation of the dual gauge group $SU({\tilde n}_c)$,
where ${\tilde n}_c = n_f-n_c$.  
The minima of the potential following from Eq.(\ref{dualqeq}) can be
found straightforwardly, and gives
\begin{equation}
  {\cal N}_2= \sum_{r=0}^{{\tilde n}_c-1} \, 
  _{n_f}\!  C_ r \, ( {\tilde n}_c-r ) 
  \label{extrav}
\end{equation} 
vacua.  The solutions have a
color-flavor diagonal form for $q$'s and ${\tilde q}$'s, with $r$
nonzero elements, $d_i, \,\, {\tilde d}_i$, where
\begin{equation}   
	d_{i} = \tilde{d}_{i} 
	= \left[ -m_{i} - \frac{1}{n_{c}+r-n_{f}} \sum_{j=r+1}^{n_{f}} m_{j}
		\right]^{1/2},
	\qquad i=1,2,\ldots,  r. 
\end{equation}
The meson vacuum expectation value
(VEV) is orthogonal to the squark VEVS,
\beq     M = {\hbox {\rm
    diag}} \, (0,0,\ldots, 0, \lambda_{r+1}, \lambda_{r+2}, \ldots,
\lambda_{n_f}), \qquad   
	\lambda_{i} = \mu \left[ m_{i} 
		+ \frac{1}{n_{c}+r-n_{f}} \sum_{j=r+1}^{n_{f}} m_{j} \right] .
	\label{vevoflam}
\end{equation}
All   VEVS    of fields carrying flavor quantum
numbers thus vanish in the limit $m_i \to 0$, showing that the flavor
symmetry remains unbroken in this class of vacua.
   
The problem is that the number of vacua found this way is too small,
since we know that the exact number of vacua is ${\cal N}$
(Eq.(\ref{nofvac})), ${\cal N} > {\cal N}_2.$ Where are other vacua?

This apparent puzzle can be solved once the nontrivial $SU({\tilde
  n}_c)\,\,$ instanton effects are taken into account
properly.\footnote{In fact, a related puzzle is how Seiberg's dual
  Lagrangian \cite{Sei} - the first two terms of Eq.~(\ref{dualqeq}) -
  can give rise to the right number of vacua for the massive $N=1$
  SQCD with $n_f > n_c+1$.  By following the same method as below but
  with $\mu = \infty$, we do find the correct number ($n_c$) of vacua.
  }  If the meson vacum expectation values have rank $n_f$, the dual
quarks can be integrated out, leaving the effective superpotential,
\begin{equation}
  W_{\it eff}  =  -{1 \o 2 \mu} \left[ \Tr M^2 - {1 \o n_c}(\Tr M)^2 
  \right]  
  + \Tr (M m )  +{\Lambda_1^{(3n_c - n_f)/(n_c-n_f)}   
    (\det M)^{1/(n_f-n_c)}}.
\label{splarmu}
\end{equation}
Minimization of this effective action gives ${\cal N}_1 = ( 2 n_c- n_f
) \cdot 2^{ n_f -1} $ solutions, having the same forms as
Eq.(\ref{missing})-Eq.(\ref{usefuleq2}).  At this point, one can make
a highly nontrivial consistency check: by changing $r \to n_f-r$ and
rearranging terms, one shows that the total number of quantum vacua is
equal to
\begin{equation}     
  {\cal N}_1 +{\cal N}_2 =\sum_{r=0}^{n_c-1}   \, (n_c-r)
  {}_{n_f} C_r = {\cal N}, 
\end{equation}
i.e., equal to the total number of semi-classical vacua.

We find therefore that there are two types of vacua: the first of
them, with finite VEVS of mesons (in $m_i \to 0$
limit), are present for all values of flavors.  They are classified by
an integer $r \leq
[n_f/2]$, and the flavor symmetry is spontaneously broken as $U(n_f) \to
U(r) \times U(n_f-r)$.  In the second type of vacua, present only for
large flavors ($n_f \ge n_f +1$), the flavor symmetry remains
unbroken.  The second type of vacua are closely related to the
emergence of the dual gauge group of Seiberg.

The analysis in the case of $USp(2n_c)$ models is similar, but the
result is qualitatively different.  We find again two types of $N=1$
vacua.  The first type of vacua has finite meson vacuum expectation values 
$M^{ij} \propto J^{ij}$ (symplectic matrix) 
with the flavor $SO(2n_f)$ symmetry broken as
\begin{equation}   
  SO(2n_f) \to U(n_f), 
  \label{SOtoU}
\end{equation}
in {\it all}\/ vacua of this class. This phenomenon is quite   reminiscent of what is believed to
occur in the standard QCD.  
The number of this type of vacua is given by
\begin{equation}  
  {\cal N}_1 =(2 \, n_c+2-n_f) \, 2^{n_f-1}, 
\end{equation}
which is the  number of vacua  for  ($n_f < n_c + 2$).  

As in the $SU(n_c)$ case, when the number of the flavor is
sufficiently large ($n_f \ge n_c + 2$) we find also another class of
vacua in which the flavor $SO(2n_f)$ symmetry is unbroken.  There are
\begin{equation}
  {\cal N}_2 = \sum_{r=0}^{n_f - n_c - 2}  \,  (n_f-n_c-1-r) \, {}_{n_f}\!C_r
\end{equation}
of them, and together with those of the first group, they make up the
total number
\begin{equation}  
  {\cal N}=    {\cal N}_1 + {\cal N}_2= \sum_{r=0}^{n_c} (n_c+1-r)  \,
  {}_{n_f}\! C_r
\end{equation}
which is the correct number of $N=1$ vacua for $USp(2n_c)$ (see
Eq.(\ref{Nvspnclass})).

\smallskip 

\noindent  {\bf  4.}       We now seek for a  microscopic  understanding of   
 the  mechanism of    dynamical flavor symmetry breaking.   We do so   by
studying the $N=2$ vacua on the Coulomb branch which survive $\mu \neq
0$ perturbation.        
We start from the auxiliary genus $n_c-1$ ($n_{c}$) curves for
$SU(n_c)$ ($USp(2 n_c)$) theories
\begin{equation}
  y^{2} = \prod_{k=1}^{n_{c}}(x-\phi_{k})^{2} + 4 \Lambda^{2n_{c}-n_{f}}
  \prod_{j=1}^{n_{f}}(x+m_{j}), \quad   SU(n_c), \, \,  n_f \le 2n_c-2,   
\end{equation}
with $\phi_{k}$ subject to the constraint $\sum_{k=1}^{n_{c}}\phi_{k} =
0$, and
\begin{equation}
  x y^{2} = \left[ x \prod_{a=1}^{n_{c}} (x-\phi_{a}^{2})^{2}
    + 2 \Lambda^{2n_{c}+2-n_{f}} m_{1} \cdots m_{n_{f}} \right]^{2}
  - 4 \Lambda^{2(2n_{c}+2-n_{f})} \prod_{i=1}^{n_{f}}(x+m_{i}^{2}),
  \quad USp(2 n_c).  \label{eq:curve}
\end{equation}
The VEVS of $a_{Di}, \,\, a_{i}$ are
constructed as  
 integrals   over the non-trivial cycles   of the  meromorphic differentials
on the  curves.
  We require that 
the curve is maximally singular, i.e. $n_c-1$ (or $n_c$ for 
$USp(2n_c)$) pairs of branch points
to coincide: this determines the possible values of $\{\phi_a\}$'s.  
These correspond to the $N=1$ vacua, with the particular $N=1$
perturbation, Eq.(\ref{adjointmass}).  Note that as we work with
generic and nonvanishing quark masses,  this is an unambiguous
procedure   to identify all the $N=1$ vacua of  our interest. 
\footnote{There are other kinds of singularities of $N=2$ QMS at
  which, for instance, three of the branch points meet.  These
  correspond to $N=1$ vacua, selected out by different types of
  perturbations such as $\Tr \Phi^3$, which are not considered here. }

We find in this way precisely the same number (${\cal N}$) of $N=1$
vacua, where ${\cal N}$ was determined earlier by the semi-classical
and large $\mu$ analyses.  In each vacuum, there are $n_c-1$  (or 
$n_c$) different kinds of
massless magnetic monopoles, corresponding to maximal Abelian subgroup
of $SU(n_c)$ or of $USp(2n_c)$.

At {\it small} generic quark masses, we observe that these
singularities group into approximate multiplets of vacua, with
multiplicities ${}_{n_f}\!C_r$, $\,r=0,1,2,\ldots, [n_f/2]$, in the
case of $SU(n_c)$, while they appear in $2^{n_f -1}$-plets plus
certain number of other vacua, in the case of $USp(2 n_c)$ theories.
Their positions are compatible with the approximate discrete symmetries, $
Z_{2n_c-n_f} $ or $ Z_{2n_c +2 - n_f}$.  We have made an extensive
numerical study in the case of rank two gauge groups with   all   possible
numbers of flavor, as well as general analytical study of these
phenomena for higher-rank groups.

As $m_i \to 0$ (or equal mass limit in the case of $SU(n_c)$) each
multiplet of vacua collapse into one multiple vacuum.  This behavior
might suggest a more or less straightforward generalization of what
occurs in $SU(2)$ gauge theories, mentioned at the beginning.  
Indeed, monopoles can     acquire nontrivial  flavor quantum numbers   as  shown by    Jackiw and   Rebbi \cite{JR} through the
fermion zero modes.  In $SU(n_{c})$ theories, by acting fermion zero mode  operators     $d_{i}$, 
$d^{\dagger}_{j}$ on the monopole state $|\Omega\rangle$, such as
\begin{equation}
 d^{\dagger}_i      | \Omega\ket,    \,\,   d^{\dagger}_{i_1}   d^{\dagger}_{i_2}  | \Omega\ket,    \,\, 
\ldots \,\,, 
  d^{\dagger}_{i_1}    \cdots  d^{\dagger}_{i_{n_f}} | \Omega\ket, \,\,
\end{equation}
we find  semi-classical monopoles belonging to anti-symmetric tensor representations 
of $U(n_{f})$.  It might appear then  the dynamical flavor symmetry breaking
(\ref{symmetrybr})  is     caused by the
condensation of  such monopoles.   As mentioned earlier, however,  this picture    would lead to far too 
many Nambu--Goldstone multiplets (except for $r=0,1$).  
The same analysis   for  $USp(2n_c)$  case shows that the semi-classical monopoles are
in the spinor representation of $SO(2n_f)$, and their condensation
would give the symmetry breaking (\ref{SOtoU}) and the number of vacua
$2^{n_f-1}$.  We  would   again run into a paradox of  having  a too-large
$SU(2^{n_f-1})$ symmetry.

Actually the theories avoid falling into this kind of paradox, but do
so in a subtle   way.  Let us discuss below physics of $SU(n_c)$ and 
$USp(2 n_c)$ gauge theories separately.

\smallskip 

\noindent  {\bf 5.}    In the ${ { SU(n_c)} }$  case, 
the $N=1$ vacua   can all be
generated from the various classes of superconformal theories with
$m_i=\mu=0$, by perturbation by masses $m_i$.  The first type   of
vacua (with multiplicity ${\cal N}_1$) correspond to the curves
\begin{equation}   
  y^2\sim   x^{2r} (x-\alpha_{1})^{2} \cdots
  	(x-\alpha_{n_{c}-r-1})^{2} (x-\beta) (x-\gamma),  
	\qquad        r=0,1,2,\ldots, [n_f/2],   
\end{equation}
that is
\begin{equation}    
  {\hbox {\rm diag}} \, \phi =(\,  {\underbrace  {0,0, \ldots, 0}_{r}},   
  \phi_1, \ldots \phi_{n_c-r} ), \qquad  
\sum_{a=1}^{n_c-r}  \phi_a=0,  
\end{equation} 
with $ \phi_a$'s chosen such that the nonzero $2(n_{c}-r-1)$ branch points are
paired.  These correspond to the so-called class 1 ($r<n_f/2$) and 3 ($r=n_f/2$, with $n_c-n_f/2$ odd)
superconformal theories  \cite{EHIY},  while  the case, $r=n_f/2$,  $n_c-n_f/2$ even,     may
be interpreted as belonging to   class 4.  Since these adjoint VEVS   break the discrete symmetry
spontaneously, they appear in
$2n_c - n_f$ copies.\footnote{There is an exception to this.  In the case of
  $r=n_f/2$ with $n_f$ even, the explicit configuration of $ \phi_a$'s
  can be found by using the Chebyshev polynomials.  This vacuum
  respects $Z_2$ subgroup of the $Z_{2n_c-n_f}$ symmetry, showing that
  it appears in $n_c - n_f/2 $ copies rather than $2n_c - n_f.$ This
  fact is crucial in the vacuum counting below Eq.(\ref{special}).}
When (generic) quark masses are turned on, these vacua split into
${}_{n_f} C_r $-plet of single vacua.
The second class of vacua stem from the (trivial) superconformal theory
\begin{equation}   
  y^2\sim   x^{2{\tilde n}_c}  
  ( x^{n_c - {\tilde n}_c } -   \Lambda^2   )^2, \qquad   
  {\tilde n}_c = n_f - n_c, 
  \label{second}
\end{equation}
corresponding to the singularity
\begin{equation}       
  {\hbox {\rm diag}} \, \phi = 
  (\, {\underbrace  {0,0, \ldots, 0}_{{\tilde n}_c}  },   
  \Lambda \,\omega,  \ldots,  \Lambda \, \omega^{n_c - {\tilde n}_c}) 
  \label{second1}
\end{equation}
with $\omega= e^{2 \pi i / (n_c - {\tilde n}_c
  )}$.
Actually there is no vacuum of the 
first type with $r=\tilde{n}_{c}$.

The most detailed description of these $N=1$ vacua comes from the
considerations based on nonrenormalization theorem of the Higgs branch
metric \cite{ArPlSei}.  The first class of vacua with given $r$ is an
$SU(r)\times U(1)^{n_c-r}$ gauge theory with $n_f$ ``quarks" and
$n_c-r$ singlet monopoles $e_k$'s, with an effective Lagrangian,
\begin{equation}
  W_{\it non bar} = \sqrt2 \Tr (q \phi {\tilde q}) + \sqrt2 \psi_0 \Tr (q 
  {\tilde q}) + \sqrt 2 \sum_{k=1}^{n_c-r-1} \psi_k e_k {\tilde e}_k + 
  \mu \left(\Lambda \sum_{i=0}^{n_c-r-1} x_i \psi_i 
  	+ {1 \o 2} \Tr \phi^2\right),
  \label{nonbaryonic} 
\end{equation}
where $\phi$ and $\psi_k$'s are part of the $SU(r)\times U(1)^{n_c-r}$ 
$N=2$ vector multiplets and $x_{i}\sim O(1)$ constants.  
These are at the roots of the so-called 
``non-baryonic" branches \cite{ArPlSei}, where they meet the Coulomb 
branch.  They describe an infrared-free (i.e., non asymptotic free) 
effective theory for $r < n_f/2$.  We now add the mass terms
\begin{equation}   
  \Tr ( m q {\tilde q})  -  \sum_{k,i} S_k^i  m_i  e_k {\tilde e}_k
  \label{masses} 
\end{equation} 
and minimize the potential.     We find ${}_{n_f} C_r $     solutions characterized by the vacuum expectation values
  ($q$ and ${\tilde q}$ having color-flavor diagonal form, with 
  nonvanishing elements, $d_i$ and ${\tilde d}_i$)\footnote{Actually, 
  Eq.(\ref{nonbaryonic}) and Eq.(\ref{masses}) allow for a number of 
  other solutions in which the vev of $\psi_0$ is of $O(\Lambda)$;
  these are the first group of $N=1$ vacua found in \cite{ArPlSei}.
  Such solutions, involving fluctuations much larger than both $m_i$ 
  and $\mu$, however, lie beyond the validity of the low-energy 
  effective Lagrangian.  They should therfore be regarded as an  artefact 
  of the approximation and must be discarded.  }
\begin{equation}  
  \psi_0= -{  1 \o \sqrt {2}\,  r}    \sum_{i=1}^r    m_i ,  
  \qquad  \psi_k =   O(m_i), 
\end{equation}
\begin{equation}        
  d_i  {\tilde d_i}   
  = -\mu \left( m_i - \frac{1}{r}\sum_{j=1}^r m_j\right) 
  - \frac{1}{\sqrt{2}\, r}\mu \Lambda x_{0}; \qquad e_k {\tilde e}_k  
  \sim   \mu \Lambda. 
\end{equation}
The multiplicity ${}_{n_f} C_r $ arises from the choice of $r$ (out of
$n_f$) quark masses used to construct the solution.  In the massless
limit we find
\begin{equation}  
  d_i =   {\tilde d_i}  \sim   \sqrt{ \mu \Lambda}, 
  \qquad i=1,2,\ldots   r:  
  \label{sur} 
\end{equation}
this leads to the correct symmetry breaking pattern, $U(n_f) \to U(r)
\times U(n_f-r)$.  

For $r=n_{f}/2$, the theory at the singularity
becomes a non-trivial superconformal theory.  
There is no description of this singularity
in terms of weakly coupled local field theory.  The monodromy around the singularity shows that
the theory is indeed   superconformal (we checked this explicitly for $n_c=3$
and $n_f=4$).   Careful
perturbation of the curve by the quark masses shows that there are
$(n_{c}-n_{f}/2) \, {}_{n_{f}}\!C_{n_{f}/2}$ vacua.\footnote{Due to some reason,
however, the naive application of the effective Lagrangian
Eq.~(\ref{nonbaryonic},\ref{masses}) gives the correct number.}

The total number of the vacua of this type   
is ($n_{f}\leq n_{c}$):
\begin{equation}    
  (2 n_c - n_f)  \sum_{r=0}^{(n_f-1)/2}  {}_{n_f} C_r  
  =   (2 n_c - n_f)  \, 2^{n_f-1}, \quad  (n_f = {\hbox {\rm odd}})   
\end{equation}
\begin{equation}     
  (2 n_c - n_f) \sum_{r=0}^{{n_f}/2-1}  {}_{n_f} C_r   
  \,   +  {2 n_c - n_f \o 2}\,  {}_{n_f} \!C_{n_f/2}   
  =   (2 n_c - n_f) \, 2^{n_f-1}, 
  \quad  (n_f = {\hbox {\rm even}}), 
  \label{special} 
\end{equation}
which      exhausts ${\cal N}$,  Eq.(\ref{nofvacbis}).  
In Eq.(\ref{special}) we have taken into 
account the fact that for even $n_f$, the vacua with $r=n_f/2$ do not 
transform under $Z_{2n_c-n_f}$ but only under $Z_{n_c-n_f/2}.  $ When 
$n_{f}>n_{c}$, we need to exclude the term  
$r=\tilde{n}_{c}=n_{f}-n_{c}$  from   the sum because it gives the second    
type of vacua.  We obtain therefore ${\cal N}_{1} - (2 n_c - n_f)\,
{}_{n_f} \!C_{\tilde{n}_{c}}$ vacua.

As for the second group of vacua, Eq.(\ref{second}),
Eq.(\ref{second1}), they are an $SU({\tilde n}_c)\times
U(1)^{n_c-{\tilde n}_c }$ gauge theory with $n_f$ ``quarks" and
$n_c-{\tilde n}_c $ singlet monopoles $e_k$'s \cite{ArPlSei}.  The
effective low-energy Lagrangian for this theory is given by
\begin{equation}  
  W_{\it bar} = \sqrt2 \Tr (q \phi {\tilde q}) 
  + {\sqrt2 \o {\tilde n}_c} \Tr (q {\tilde q}) 
  \left( \sum_{k=1}^{n_c - {\tilde n}_c} \psi_k \right) 
  - \sqrt 2 \sum_{k=1}^{n_c-{\tilde n}_c} \psi_k e_k {\tilde e}_k 
  + \mu \left(\Lambda \sum_{i=1}^{n_c-{\tilde n}_c} x_i \psi_i 
  + {1 \o 2} \Tr \phi^2\right) \, .
  \label{baryonic} 
\end{equation}
where $\phi$ and $\psi_k$'s are now in $SU({\tilde n}_c)\times
U(1)^{n_c-{\tilde n}_c}$ $N=2$ vector multiplets.  We add the mass
terms
\begin{equation} 
  \Tr ( m q) - \sum_{k,i} S_k^i m_i e_k {\tilde e}_k.
  \label{baryomas}
\end{equation}
We find two types of vacua of the effective low-energy Lagrangian.
The first type has $e_{k}=\tilde{e}_{k} = (\mu \Lambda
x_{k}/\sqrt{2})^{1/2}$ for all $k=1, \cdots n_c-\tilde{n}_c$.  
Minimizing the potential in this case, we find
\begin{equation}
  {\cal N}_2   = \sum_{r=0}^{\tilde n_c-1} ( \tilde n_c-r) \,\, 
  {}_{n_f}\!C_r 
  \label{nmvcbar}
\end{equation}
$N=1$ vacua, characterized by the vevs
\begin{equation}
  \phi = \frac{1}{\sqrt{2}}\diag \, (-m_1, \ldots, -m_r, c, \ldots , c) 
  \, ; \qquad c = {1 \o {\tilde n}_c-r} \sum_{k=1}^r m_k \, 
  \label{diagphi2}
\end{equation}
\begin{equation}  
  d_i ,   {\tilde d_i}   \sim  \sqrt {\mu    m} 
  \stackrel{ m_i \to 0}{\longrightarrow}   0, 
  \qquad \qquad e_k,   {\tilde e}_k  \sim  \sqrt{\mu\Lambda}.  
\end{equation}  
The unbroken $SU(\tilde{n}_{c}-r)$ gauge group gives 
$\tilde{n}_{c}-r$ vacua each.
These vacua describe the vacua with unbroken $SU(n_f)$ symmetry,  which  are known to exist from the
large $\mu$ analysis.  

The second type of vacua in Eqs.~(\ref{baryonic},\ref{baryomas}) has 
one of the $e_{k}=\tilde{e}_{k}=0$ (hence $n_c-{\tilde 
n}_c=2n_{c}-n_{f}$ choices) while $\partial W/\partial\psi_{k}=0$ 
requires quarks to condense with $q = \tilde{q} \sim \sqrt{\mu 
\Lambda}$.  Dropping $e_{k}=\tilde{e}_{k}=0$ from the Lagrangian, it 
becomes the same as that of the non-baryonic root  
Eqs.~(\ref{nonbaryonic},\ref{masses}) and gives 
$(2n_{c}-n_{f}) \, {}_{n_{f}}\!C_{\tilde{n}_{c}}$ vacua.  This precisely 
compensates the exclusion of $r=\tilde{n}_{c}$ in the sum for the 
non-baryonic roots and the correct total number of vacua ${\cal 
N}_{1}+{\cal N}_{2}$ is obtained.

We thus find that both the number and the symmetry properties of the
$N=1$ theories at small adjoint mass $\mu$ exactly match those found
at large $\mu,$   without encountering  any  paradoxical situation.     We postpone a discussion of physical aspects of $SU(n_c)$
theories to the end (point 7 below).

\smallskip

\noindent  {\bf 6. }   In the ${   {USp(2 n_c)} }$  
 gauge theories, the first type of vacua can be
identified more easily by first considering the equal but novanishing
quark masses.  The adjoint vevs in the curve Eq.(\ref{eq:curve}) can
be chosen so as to factor out the behavior
\begin{equation}   
  y^2 = (x+m^2)^{2r}\, [\ldots],    \quad       r=1,2,\ldots   
\end{equation} 
which describes an $SU(r)\times U(1)$ gauge theory with $n_f$ quarks.  These
(trivial) superconformal theories belong in fact to the same universality
classes as in the $SU(n_c)$ gauge theory as pointed out by
\cite{EHIY}.  They are therefore described by exactly the same
Lagrangian Eq.(\ref{nonbaryonic}).  At each vacuum with $r$, the
symmetry (of equal mass theory, $U(n_f)$) is broken spontaneously as
\begin{equation}   
  U(n_f) \to U(r) \times U(n_f-r) :  
  \label{funny} 
\end{equation}
as in Eq.(\ref{symmetrybr}).  When a small mass splitting is added   
among $m_i$'s, each of the $r$ vacuum split into ${}_{n_f}\!C_r$
vacua, leading to the total of
\begin{equation}    
  (2 n_c +2 - n_f)  \sum_{r=0}^{(n_f-1)/2}  {}_{n_f} C_r  
  =   (2 n_c +2 - n_f)  \, 2^{n_f-1}, \quad  (n_f = {\hbox {\rm odd}})   
\end{equation}
\begin{equation}     
  (2 n_c +2 - n_f) \sum_{r=0}^{{n_f}/2-1}  {}_{n_f} C_r   
  \,   +  {2 n_c +2 - n_f \o 2}\,  {}_{n_f} \!C_{n_f/2}   
  =   (2 n_c +2 - n_f) \, 2^{n_f-1}, 
  \quad  (n_f = {\hbox {\rm even}}), 
\end{equation}
vacua of this type,      consistently    with Eq.~(\ref{Nvspnclassbis}).%
\footnote{These $N=1$ vacua seem to have been
  overlooked in \cite{APS2}.}

In the massless limit the underlying theories possess a larger, flavor
$SO(2n_f)$ symmetry.  We know also from the large $\mu$ analysis that
in the first group    of vacua (with finite vevs), this symmetry is
broken spontaneously to $U(n_f)$ symmetry always.  How can such a
result be consistent with Eq.(\ref{funny}) of equal (but nonvanishing)
mass theory?

What happens is that in the massless limit various $N=1$ vacua with
different symmetry properties Eq.(\ref{funny}) (plus eventually other
singularities) coalesce.  The location of this singularity
can be obtained exactly in terms of Chebyshev polynomials.  At the singularity
there are   mutually non-local dyons and hence the theory is at a non-trivial infrared
fixed point (in the example of $USp(4)$ theory with $n_f=4$, we have
explicitly verified this by determining the singularities and branch
points at finite equal mass $m$ and by studying the limit $m \to 0$.)
There is no description
in terms of a weakly coupled local field theory, just as in     the case
$r=n_f/2$ for $SU(n_c)$ theories.  
Since the global flavor symmetry is $SO(2n_f)$ these superconformal
theories belong to different universality classes as compared to those
at finite mass.  We find this behavior resonable because the
semi-classical monopoles are in the spinor representation of the
$SO(2n_f)$ flavor group and, in contrast to the situation in $SU(n_c)$  theories,    cannot ``break up'' into quarks in the
vector representation.     They  are  therefore  likely to  persist at the
singularity and makes the theory superconformal.  Once the quark
masses are turned on, however, the flavor group reduces to (at least)
$U(n_f)$ and it becomes possible for monopoles to break up into
quarks; this explains the behavior in the equal mass case.

As for the second group of vacua, the situation is more analogous to
the case of $SU(n_c)$ theories.  The superpotential reads in this case
(by adding mass terms to Eq.(5.10) of \cite{APS2}): 
\begin{eqnarray} 
  W &=& \mu \,
  \left(\Tr \, \phi^2 + \Lambda \sum_{a = 1}^{2 n_c+2-n_f} x_a 
  \psi_a\right) 
  + { 1  \o \sqrt2} \, q_a^i \phi_b^a q_c^i \,
  J^{bc} + {m_{ij} \o 2} q_a^i q_b^j \, J^{ab}    \nonumber \\
  &+& \sum_{a = 1}^{2 n_c+2-n_f} \left(\psi_a \, e_a \, {\tilde e}_a + 
  S_a^i m_i \, e_a \, {\tilde e}_a \right)\, ,
\label{uspsuperpot}
\end{eqnarray}
where $J = i\si_2 \otimes {{\bf 1}}_{n_c}$   and
\begin{equation}
  m = - i \si_2 \otimes \diag \, (m_1, m_2, \ldots, m_{n_f}) \, .
\end{equation}
By minimizing the potential, we find
\begin{equation}
  {\cal N}_2 =   \sum_{r=0}^{{\tilde n}_c}  ({\tilde n}_c- r +1)
  {}_{n_f} C_{r}\, 
\end{equation}
vacua, which precisely matches the number of the vacua of the second
group, with squark vevvs behaving as
\begin{equation}    
  q_i, {\tilde q}_i \sim \sqrt {\mu m_i}
  \stackrel{ m_i \to 0}{\longrightarrow}   0.    
\end{equation}  
These are the desired   $SO(2n_f)$  symmetric vacua.  

\smallskip

\noindent  {\bf  7.}  To summarize, 
we  have  studied the dynamics of $N=1$ $SU(n_{c})$ and $USp(2n_{c})$ gauge 
theories obtained by perturbing $N=2$ theories with $n_{f}$ 
hypermultiplets in the fundamental representation with a finite adjoint 
mass $\mu {\rm Tr}\Phi^{2}$,  determining   the possible flavor 
symmetry breaking patterns.    There are 
vacua in confinement phase   with symmetry breaking $U(n_{f})\rightarrow U(r) \times U(n_{f}-r)$ 
($r=0, 1, \cdots, [n_{f}/2]$) and $SO(2n_{f}) \rightarrow U(n_{f})$, 
respectively.  There also  are  non-confining vacua with no flavor 
symmetry breaking for $n_{f} \geq n_{c}+1$, $n_{f} \geq n_{c}+2\,$   for  $SU(n_{c})$ and $USp(2n_{c})$ 
theories,  respectively.

With small but generic quark masses, the  order  parameter of  confining vacua    is  indeed the   
condensation of magnetic monopoles for every $U(1)$ factor on 
the Coulomb branch \`a la 't Hooft,  in both types of gauge theories.  
 The massless limit, however, is 
non-trivial and   much    more   interesting.

In $SU(n_{c})$ theories, in  vacua with $r=1$   magnetic monopoles  are  in 
the fundamental representation of $U(n_{f})$ flavor group,  and are  charged 
under one of the $U(1)$'s:   flavor-singlet monopoles  are charged under 
other $U(1)$'s.  Their condensation realizes    the confinement and the 
flavor symmetry breaking at the same time.  

In vacua   labelled by $r$,  $2\leq r < n_{f}/2$ 
but $r \neq n_{f} - n_{c}$,   the grouping of the associated singularities  on the Coulomb 
branch might suggest the condensation of monopoles in the rank-$r$ 
anti-symmetric tensor representation.    Actually,  this does not occur.    The correctness of  the effective action
Eq.(\ref{nonbaryonic})  shows that  the low-energy degrees of freedom of   these theories are  (magnetic) quarks
 plus a  number of singlet monopoles  of  an  effective      $SU(r)\times  U(1)^{n_c-r}$  gauge theory.  
Monopoles in a higher representation of $SU(n_f)$ flavor group    probably  exist    
semi-classically  as  seen in a Jackiw--Rebbi type   analysis \cite{JR}.  
Such monopoles can be interpreted    as  ``baryons''  made  of  the 
magnetic quarks,  which,  interactions being   infrared-free,   break up before they become massless at 
singularities on the Coulomb branch.       The condensation   of
the   magnetic  quarks induces     the confinement and  flavor symmetry breaking, 
$U(n_f) \rightarrow U(r) \times U(n_f-r)$,    at the 
same time.  This is how the system avoids falling into a paradox of 
having too many Nambu-Goldstone multiplets. 

 In the special cases with 
$r= n_f/2$, the interactions among the monopoles are so strong that 
the low-energy theory describing them is a nontrivial superconformal 
theory (conformal invariance explicitly broken by the adjoint or quark 
masses).  Although the symmetry breaking pattern is known ($U(n_{f}) 
\rightarrow U(n_{f}/2) \times U(n_{f}/2)$), the low energy degrees of 
freedom are fields     whose interactions are not described by a local 
action.  

Finally,  in the group of vacua labelled by  $r=n_{f}-n_{c}$, the interactions among 
monopoles are described by an effective  
infrared-free $SU(n_f-n_c)$   gauge theory.  There are two physically distinct groups    of vacua in this case: 
one in which   the magnetic quarks   condense   (i.e.   confinement phase)   with the unbroken 
symmetry $U(n_{f}-n_{c}) \times U(n_{c})$, and the other with  no
magnetic-quark condensation  and hence with  unbroken $U(n_{f})$ symmetry  (i.e.  the free magnetic phase).  

In $USp(2n_c)$ theories,      physics   for non-vanishing and  equal quark masses   resembles  that in  the 
vacua with generic $r$  of $SU(n_c)$  theory.  In the
massless limit, however,     where the flavor group enlarges to $SO(2n_{f})$, the situation is more 
similar to the $r=n_f/2$ case of $SU(n_c)$: the 
low-energy degrees of freedom are fields   with relatively non-local 
interactions, and the effective theory is a non-trivial superconformal 
one.  Although no local effective Lagrangian is available, we   know 
that the flavor $SO(2n_f)$ symmetry is spontaneously broken to 
diagonal $U(n_f)$ symmetry in all confining vacua.  For large number of flavor,  there are 
also     vacua in free-magnetic phase.

A more extensive account of our analysis will appear elesewhere.

\newpage  

\noindent{\bf Acknowledgment}

One of the authors (K.K.) thanks Lawrence Berkeley National
Laboratory, University   of California, Berkeley, and ITP, University of
California Santa Barbara, for their   warm hospitalities.  Part of the work was done during the workshop,
``Supersymmetric Gauge Dynamics and String Theory"  at ITP,  UCSB   to which  two of us (G.C. and K.K.) 
participated.     The authors
acknowledge useful discussions with Philip Argyres, Prem Kumar, Misha
Shifman and Arkady Vainshtein.  This research was supported in part by
the National Science Foundation under Grant No. PHY-94-07194,
PHY-95-14797, and in part by the Director, Office of Science, Office
of High Energy and Nuclear Physics, Division of High Energy Physics of
the U.S. Department of Energy under Contract DE-AC03-76SF00098.

\end {document}